# Development of a Quasi-Periodic Undulator for the HLS

Yang Yu-feng （杨宇峰）[1], Lu Hui-hua（陆辉华）[1,*], Chen Wan （陈宛）[1], Jia Qi-ka （贾启卡）[2], Sun Shu-chen （孙舒晨）[1], Li Zhi-qiang （李志强）[1]

1. Institute of High Energy Physics, Chinese Academy of Sciences, 100049, Beijing, China

2. University of Science and Technology of China, 230029, Hefei, Anhui, China

* Corresponding author

**Abstract**：China's first quasi-periodic undulator (QPU) has been developed for the Hefei Light Source (HLS). It uses a magnetic configuration with varied thicknesses of NdFeB blocks, based on the QPU of European Synchrotron Radiation Facility (ESRF). Depression of 3rd harmonic radiation is significantly improved over the ESRF QPU, as deduced from the measured magnetic fields. A method of configuring shims of different geometries and sizes, based on a symmetric principle to correct multi-pole field integrals, was demonstrated.

**Keywords**：quasi-periodic undulator, high harmonic radiations, multi-pole field integrals

**PACS**：29.20.db, 07.85.Qe

## 1. Introduction

China's first quasi-periodic undulator (QPU) was planned for the 800 MeV electron storage ring of the Hefei Light Source (HLS). Users hoped the 3rd, 5th and 7th harmonic radiations could be greatly depressed.

Schematically, a QPU may be realized by introducing a phase slip in a periodic undulator[1]. Several types of magnetic configuration have been proposed,[2-5] with the most compact scheme coming from ESRF [5]. The ESRF design modifies a Halbach pure permanent magnet type undulator by vertically retracting some H-magnets to interrupt the periodicity of the undulator in an irrational way. As a result, rational harmonic radiations are replaced by irrational ones in the radiation spectrum.

To design the HLS QPU, we changed the longitudinal thickness ratio of two adjacent magnetic blocks in the ESRF QPU. As indicated by recent numerical studies [6-9], harmonic fields and radiations may be optimized in this way.

In this paper, we describe the development of the HLS QPU with regard to magnetic design, manufacturing, commissioning and magnetic field tests. High harmonic radiations properties deduced from the measured fields were compared with the ESRF QPU.

To meet the beam dynamics requirements of a storage ring, the method of symmetric configuration of shims, i.e., thin iron pieces, of different geometries and sizes to correct multi-pole field integrals is investigated.

## 2. Magnetic Design

The HLS QPU is a 1.7 m-long gap-adjustable undulator with 19 periods and 88 mm period length. It is a pure permanent magnet made of NdFeB. The arrangement of magnetic blocks along the beam direction is V-H-V-H..., as Fig.1 shows, where V and H refer to blocks with vertical and horizontal magnetization respectively. Similar to the ESRF QPU, some H-blocks are vertically displaced by some distance $\delta$ and an irrational dimensionless number $\eta$ called the inter-lattice ratio is used to decide which H-blocks will be moved [5].

Let $L_V$ and $L_H$ be the thicknesses of the V-blocks and H-blocks respectively in the ESRF QPU, $L_V = L_H = \lambda_u / 4$, where $\lambda_u$ is the period length. The HLS QPU differs from the ESRF QPU in that it makes $L_V \neq L_H$ to optimize harmonic radiations. After optimization, the parameters of the HLS QPU are set to be $\eta = \sqrt{5}$, $\delta = 5mm$ and $L_V / L_H = 14:30$. According to simulations, this gives an improvement over $L_V / L_H = 1:1$ in depressing the 3rd and 7th harmonics (Tab.1).

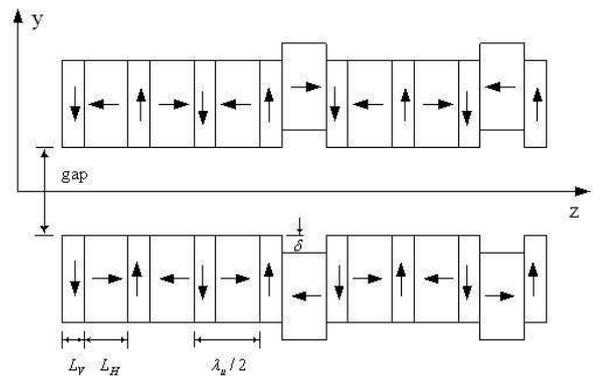

Fig.1 Schematic of HLS QPU

| $L_V / L_H$ | Fundamental energy [eV] | Fundamental flux density [ph/s/mrad$^2$/0.1%] | Harmonics flux normalized to fundamental | | |
|---|---|---|---|---|---|
| | | | 3rd | 5th | 7th |
| 22:22[A] | 4.65 | 0.68E15 | 0.435 | 0.013 | 0.125 |
| 14:30[A] | 5.12 | 0.68E15 | 0.087 | 0.204 | 0.118 |
| 14:30[B] | 4.77 | 5.8E+14 | 0.026 | 0.151 | 0.080 |

Tab.1 Radiations computed with SPECTRA [10], based on magnetic fields from: RADIA[11] simulations (superscript A) and experimental measurements (superscript B) . Parameters: $\lambda_u = 88mm$, Gap=32mm, $\eta = \sqrt{5}$, $\delta = 5mm$.

## 3. Manufacturing

Construction of the HLS QPU includes manufacturing and assembly of the magnetic materials, the supporting structure and the control system. The magnetic materials of the NdFeB blocks were sorted before being mounted on the girders to reduce field errors induced by their individual defects. Mechanical and control accuracy tests were done at each step of the process. The complete device is shown in Fig.2.

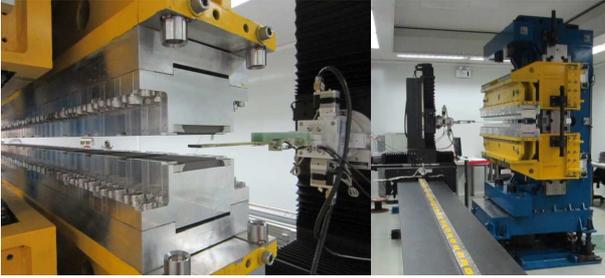

Fig.2 The HLS QPU set up

## 4. Field Correction

Though special attention is paid to reducing field errors, it is impossible to totally eliminate them in manufacturing. Further correction is needed to meet the beam dynamics requirements. There are several tools available, such as correction coils and magic fingers, which can achieve field correction. However, their installation requires additional space which is not available in the HLS. Therefore thin iron pieces called shims were applied. The advantage of shims over the previous tools is that they can be installed anywhere in an undulator. This makes them suitable for correcting field errors on the spot, which brings a lot of convenience to error elimination.

In this section, a method to correct multi-pole field integrals using shims is investigated and applied to the HLS QPU. The shimming method of on-axis field integrals and their gap-dependence has been studied in [12-14].

Off-axis field integrals $I_1(x)$ are involved in the evaluation of multi-pole field integrals. By polynomial fitting of $I_1(x) = \sum_n A_n x^n$, the multinomial coefficient $A_n$ is derived corresponding to the 2(n+1)$^{th}$ order of multi-pole field integrals, which exist only when $I_1(x) \neq 0$ and vanish when $I_1(x) = 0$. An approach to eliminate multi-pole field integrals is to make $I_1(x) = 0$, which may be realized by combining shims of different geometries and sizes which induce additional $I_1(x)$ signatures matching the existing error.

As verified by numerical simulations and experimental measurements, the $I_1(x)$ induced by thin shims, e.g., less than 0.5mm, follows the superposition principle which means the contribution of a combination of several shims equals the sum of the individual shims. On this basis, given the contribution to $I_1(x)$ of each single shim, the total contributions to $I_1(x)$ of the combined shims are known. To find a solution, one needs to try many different combinations of shims with different geometries.

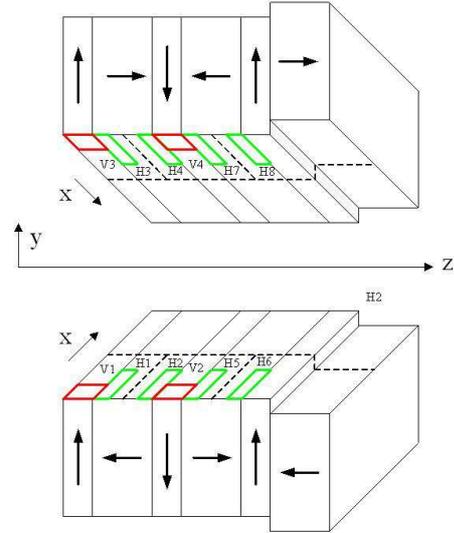

Fig.3 Diagram of equivalent positions of V-block shimming and H-block shimming. Red and green rectangles represent shims.

|            | V1 | V2 | V3 | V4 |
|------------|----|----|----|----|
| $I_{1x}(x)$ | A  | -A | -A | A  |
| $I_{1y}(x)$ | B  | -B | B  | -B |

|            | H1 | H2 | H3 | H4 | H5 | H6 | H7 | H8 |
|------------|----|----|----|----|----|----|----|----|
| $I_{1x}(x)$ | C  | -C | -C | C  | C  | -C | -C | C  |
| $I_{1y}(x)$ | D  | -D | D  | -D | -D | D  | -D | D  |

Tab.2 Signs principle for $I_1(x)$ induced by a shim on equivalent positions. V1-V4 and H1-H8 are shown in Fig.3.

The $I_1(x)$ induced by a rectangular shim depends on its size and the position where it is installed. For a specific sized shim, there are different positions that offer equal or negative $I_1(x)$ signatures, as shown schematically in Fig.2. The positions with equal $|I_1(x)|$ are called equivalent positions. If the shimming position in question is not located on or adjoining the displaced H-block, there are four equivalent positions in each period for V-block shimming (V1-V4 in Fig.2) and eight for H-block shimming (H1-H8 in Fig.2). The signs principle for $I_1(x)$ is summarized in Tab.2. Note that equivalent positions are always distributed on the same x side with no horizontal offset.

On this basis, one can easily create a corrective $I_{1x}(x)$ with $I_{1y}(x)=0$ or $I_{1y}(x)$ with $I_{1x}(x)=0$ by combining two identical shims at appropriate equivalent positions, making it possible to independently correct $I_{1x}(x)$ and $I_{1y}(x)$ with no mutual effect.

## 5. Magnetic Field Tests

Magnetic measurements were made using the bench at the Institute of High Energy Physics (IHEP)[15]. The vertical $B_y$ field was measured using a Hall probe with a sensitive area of about 1 mm in diameter with residual error in the calibration of $2\times10^{-5}$ T. The horizontal field $B_x$ was measured using a sensor coil of dimensions $3.95\times13.55\times7.0$ mm$^3$ with 7000 turns and a calibrated winding area of about 0.27 m$^2$.

After applying shim and pole height tuning, the magnetic fields and 2$^{nd}$ field integrals are shown in Fig.4. The 1$^{st}$ field integrals $|I_1(x)|<100 Gscm$ and 2$^{nd}$ field integrals $|I_2(x)|<20000 Gscm^2$ sit within the HLS specifications. Multi-pole field integrals are reduced to acceptable levels, as Fig.5 shows, where normal and skew quadrupole integrals are 8.8Gs and 22.6Gs, sextupole integrals are 5.1Gs/cm and 12.2Gs/cm, and octupole integrals are -9.6 Gs/cm$^2$ and -16.2 Gs/cm$^2$ at gap=32mm.

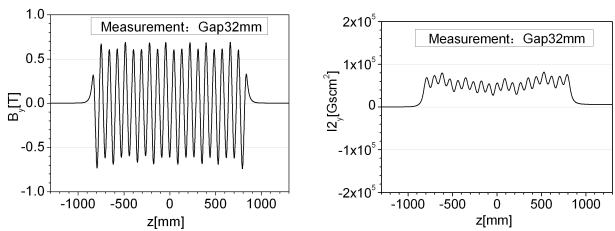

Fig.4 On-axis fields (left) and 2$^{nd}$ field integrals in the vertical direction at gap=32mm after shimming.

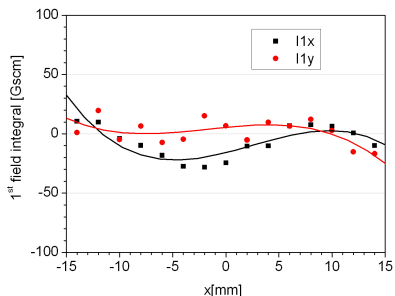

Fig. 5  1$^{st}$ field integral along x direction. Dotted line: measured data; smooth line: polynomial fit to the measured data.

The radiation spectrum was computed using SPECTRA [10] based on the measured fields, as shown in Fig.6 and Tab.1. There is a fundamental deviation between the energy deduced from RADIA [11] and the measured fields (Tab.1), attributed mainly to the conservative estimate of NdFeB remanence used in RADIA. High harmonic radiations are significantly depressed, where the 3$^{rd}$ harmonic is depressed by 58.5 times at gap=32mm by taking into account the inherent difference of each harmonic. In comparison, the same figure for the ESRF QPU with uniform-sized magnets is 8.3[5].

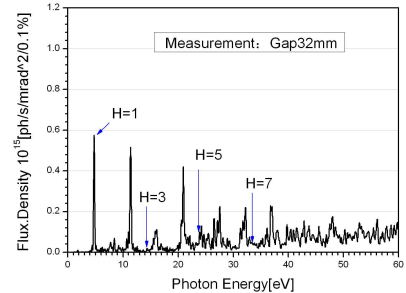

Fig.6    Radiation spectrum computed using measured on-axis fields

## 6. Conclusions

China's first QPU was developed for the HLS by changing the thickness ratio of adjacent magnetic blocks, using a design based on the ESRF QPU. High harmonic radiations are significantly depressed compared to the current QPU as deduced from the measured magnetic fields.

A symmetric method to configure shims of different geometries and sizes was developed for multi-pole field correction. The symmetry principle for multi-pole field correction is stricter than that for on-axis field correction [14-15] because the horizontal offset must be taken into account. In this way, horizontal and vertical magnetic field errors can be corrected independently without any mutual interference.

A significant advantage of shimming over magic fingers and correction coils is that shims can be installed anywhere and give a site-specific correction. As a result, errors are cured on the spot and exert no further impact on beam dynamics. Its disadvantage for surface flatness is not entirely a problem for out-vacuum insertion devices where the electron beam passes through a vacuum chamber.

**REFERENCES**

1. M .Takao, T.Shimada, Y.Miyahara, et al., Proc. of EPAC96, ,1996, 2546;
2. Sasaki S. ,Proc. of IPAC10,  2010, 3141;
3. M .Kawai, M .Yokoyama et al., Proc. of EPAC96,.1996, 2549.
4. S. Hashimoto, S.Sasaki, Nucl. Instr. And Methods,1995, A361. 611;
5. J .Chavanne, P.Elleaume P, et al, Proc. of EPAC98, 1998, 2213;
6.  Jia Qika, Phys. Rev. ST  Accel. Beams:,2011,14 (6),060702;
7. Jia Qika, Wu Ailin , Nucl. Instr. And Methods, A685, ,2012, 7;

8. Jia Qika, Proc. of IPAC2013,,2013, ,2177;
9. Jia Qika, Proc. of IPAC10, 2010, 3165 ;
10. SPECTRA Code, http://radiant.harima.riken.go.jp/spectra/;
11. RADIA code, http://www.esrf.eu/Accelerators/Groups/InsertionDevices/Software/Radia;
12. Li Yuhui and Joachim Pflueger, Internal XFEL.EU report WP71/2012/15;
13. Yang Yufeng, Li Yuhui, Lu Huihua, et al, Internal XFEL.EU report WP71/2013/04;
14. Yang Yufeng, Li Yuhui, Lu, Huihua, et al,, Proc. of FEL2013, 2013, 313;
15. Lu Huihua, Chen Wan, et al. Proc. of FEL2011, 2011, 420.